\documentclass[12pt]{iopart}
\usepackage{graphicx}
\usepackage{color}
\newtheorem{proposition}{Proposition}
\newtheorem{lemma}{Lemma}

\begin{document}
	\title[Effective Gibbs state for averaged observables]{Effective Gibbs state for averaged observables}
	\author{Alexander E Teretenkov}
	\address{Department of Mathematical Methods for Quantum Technologies,\\ Steklov Mathematical Institute of Russian Academy of Sciences,\\ ul.\,Gubkina 8, Moscow 119991, Russia.}
	\ead{\mailto{taemsu@mail.ru}}

	\begin{abstract}
		We consider the effective Gibbs state for averaged observables. In particular, we perturbatively calculate the correspondent effective Hamiltonian. We show that there are a lot of similarities between this effective Hamiltonian and the mean force Hamiltonian. We also discuss a thermodynamic role of the information loss due to restriction of our measurement capabilities to such averaged observables. 
	\end{abstract}


		\section{Introduction}
	
	There are a lot of physical models, which use averaging with respect to fast oscillations  one way or another. For example, many derivations of master equations use secular approximation directly \cite[subsection 3.3.1]{breuerTheoryOpenQuantum2007}, \cite[section 5.2]{rivasOpenQuantumSystems2012} or as result \cite{daviesMarkovianMasterEquations1974a, accardiQuantumTheoryIts2010} of perturbation theory with Bogolubov-van Hove scaling \cite{bogoliubovProblemsDynamicalTheory1946, vanhoveEnergyCorrectionsPersistent1955} (see also corrections beyond the zeroth order in \cite{teretenkovNonperturbativeEffectsCorrections2021}). Also there is wide discussion of applicability of rotating wave approximation (RWA)  and systematic perturbative corrections to it in the literature \cite{flemingRotatingwaveApproximationConsistency2010, benattiEntanglingTwoUnequal2010, maEntanglementDynamicsTwo2012, wangEffectiveHamiltonianJaynes2021, trubilkoTheoryRelaxationPumping2020,soliverezGeneralTheoryEffective1981, thimmelRotatingWaveApproximation1999, chenSolutionsJaynesCummingsModel2011, zeuchExactRotatingWave2020a, milaStrongCouplingExpansionEffective2011}. But in this work we consider such averaging not as approximation but as restriction of our observation capabilities. And we analyze the thermodynamic equilibrium properties of a quantum system assuming such restrictions.   Due to this averaging the thermodynamic equilibrium properties can be defined by  some effective Gibbs state, which is averaged with respect to these fast oscillations, instead of the exact Gibbs state. Similarly to strong coupling thermodynamics this effective Gibbs state can be defined by some effective temperature dependent Hamiltonian, which is an analog of the mean force Hamiltonain (see e.g. \cite[chapter 22]{binderThermodynamicsQuantumRegime2018}, \cite{talknerColloquiumStatisticalMechanics2020a}, \cite{trushechkinOpenQuantumSystem2021} for recent reviews). 
	
	In section \ref{sec:effHam} we describe the setup of our problem and develop a systematic perturbative calculation for the effective Hamiltonian.    We show that the zeroth and the first term of the expansion  coincide with  the RWA Hamiltonian and, in particular, are temperature-independent. In this point it is similar to effective Hamiltonians also arising as corrections to RWA but in dynamical and non-equilibrium problems. The second order term is temperature-dependent. We show that  both this term and its derivative with respect to inverse temperature are non-positive definite.
	
	In section \ref{sec:tempDepEff} we show that this definiteness is closely related to positivity of the information loss  due to the fact that we have access only to the averaged observables discussed above rather than all possible observables. We show that information loss leads to energy loss, which is hidden from our observation. We  prove  (without perturbation theory) that these losses are always non-negative, but in the leading order they are defined by the second order temperature-dependent term in the effective Hamiltonian expansion. Additionally, we prove that exact non-equilibrium free energy  is always larger than the free energy  observable  in our setup. If one assumes that the effective Gibbs state is an exact state, then this difference is also defined by the second order term of the effective Hamiltonian expansion. At the end of section \ref{sec:tempDepEff} we argue that the analogy between our effective Hamiltonian and the mean force Hamiltonian is because they are special cases of the general setup, based on so-called conditional expectations.
	
	To dwell on  this analogy, in section \ref{sec:meanForce} we consider a compound system and the mean force Hamiltonian of one of the subsystems for the effective Gibbs state discussed above. We also give systematic perturbative expansion for it.
	
	In section \ref{sec:examples} we consider several simple examples to illustrate the results of the previous sections. Namely, we consider two interacting two-level systems, two interacting oscillators and a two-level system interacting with the oscillator. We calculate the effective Hamiltonians for such systems and the information losses due to restriction to the averaged observables.
	
	Both the effective Hamiltonian we define in this work and explicit perturbative expansion for it are novel, but such a Hamiltonian has much in common with the mean force Hamiltonian (see the end of section \ref{sec:tempDepEff} for more precise discussion). The main difference consists in the choice of projector. Thus, our results suggest the possibility to generalize equilibrium quantum thermodynamics to effective equilibrium quantum thermodynamics by different choices of projector.
	
	\section{Effective Hamiltonian}\label{sec:effHam}
	
	We are interested in equilibrium properties of fast oscillating  observables which are in the resonance with the free Hamiltonian. We assume that the equilibrium state has the Gibbs form
	\begin{equation}
		\rho_{\beta}  = \frac{e^{- \beta H}}{Z}
	\end{equation}
	with inverse temperature $ \beta > 0 $ and the Hamiltonian of the form
	\begin{equation}
		H = H_0 + \lambda H_I,
	\end{equation}
	where $ H_0 $ is a free Hamiltonian and $ H_I $ is an interaction Hamiltonian, $ \lambda $ is a small parameter.

	And we consider the observables which are explicitly time-dependent with very specific time-dependence. Namely, they depend on time in the Schrödinger picture as follows
	\begin{equation}
		X(t) = e^{-i H_0 t} X e^{i H_0 t}
	\end{equation}
	i.e. they depend on time in such a way, that they become constant in the interaction picture for the ''free'' Hamiltonian $ H_0 $. A widely used example of such an observable is a dipole operator interacting with the classical electromagnetic field in resonance with free Hamiltonian (see, e.g., \cite[ Sec. 15.3.1]{mandel1995optical}). And we assume that one could actually observe the long-time averages 
	\begin{equation}\label{eq:aver}
		\langle X(t) \rangle_{\rm av} = \lim\limits_{T \rightarrow + \infty} \frac{1}{T}\int_0^T \langle X(t) \rangle dt,
	\end{equation}
	where $ \langle X(t) \rangle \equiv \Tr \rho_{\beta} X(t) $. By ''long'' we mean long with respect to inverses of non-zero Bohr frequencies, where Bohr frequencies are the eigenvalues of the superoperator $ [H_0, \; \cdot \;] $ (see e.g \cite[p.~122]{accardiQuantumTheoryIts2010}). The observation of such a  long-time averages is usual for spectroscopy set-ups \cite[Sec. 4]{mukamel1995principles}.  Moreover, further we will discuss the perturbation theory in $ \lambda $ assuming that this averaging is already done, so this long timescale remains ''long'' even being multiplied by any power of $ \lambda $. Otherwise, one should introduce the small parameter  in the averaging procedure as well, which leads to more complicated perturbation theory depending on how the small parameter in the averaging and in the Hamiltonian are related to each other.
	
	Average \eref{eq:aver} can be represented as 
	\begin{equation}
		\langle X(t) \rangle_{\rm av} =  \Tr  X  \tilde{\rho}_{\beta},	
	\end{equation}
	where $ \tilde{\rho}_{\beta} $ is some effective Gibbs state, which could be calculated as
	\begin{equation}
		\tilde{\rho}_{\beta}  =  \mathcal{P} \rho_{\beta},
	\end{equation}
	where 
	\begin{equation}
		\mathcal{P}  X = \lim\limits_{T \rightarrow + \infty} \frac{1}{T}\int_0^T e^{i H_0 t} X e^{-i H_0 t} dt,
	\end{equation}
	since
	\begin{eqnarray}\fl
		\langle X(t) \rangle_{\rm av} =  \lim\limits_{T \rightarrow + \infty} \frac{1}{T}\int_0^T \Tr e^{-i H_0 t} X e^{i H_0 t} \rho_{\beta} dt  \nonumber\\
		= \lim\limits_{T \rightarrow + \infty} \frac{1}{T}\int_0^T \Tr X e^{i H_0 t} \rho_{\beta}  e^{-i H_0 t} dt = \Tr X \mathcal{P} \rho_{\beta}.
	\end{eqnarray}
	
	From the thermodynamcal point of view it is natural  to represent this  effective Gibbs state in the Gibbs-like form
	\begin{equation}
		\tilde{\rho}_{\beta} = \frac{1}{Z} e^{- \beta \tilde{H}}
	\end{equation}
	with some effective Hamiltonian $ \tilde{H} $  similarly to the mean force Hamiltonian \cite[chapter 22]{binderThermodynamicsQuantumRegime2018}. Let us remark that we have the same partition function both for exact and effective Hamiltonians due to the fact that $ \mathcal{P} $ is a trace-preserving map (see~\ref{app:avProj}) $ \Tr e^{- \beta \tilde{H}} =  \Tr  \mathcal{P}  e^{- \beta H} =  \Tr e^{- \beta H}$. Let us summarize several properties of the superoperator $ \mathcal{P} $ which will be used further (see~\ref{app:avProj} for the proof).
	
	\begin{enumerate}
		\item\label{it:complPos}  $ \mathcal{P} $ is completely positive.
		\item\label{it:proj}  $ \mathcal{P} $ is self-adjoint (with respect to trace scalar product $ \Tr X^{\dagger} Y $) projector
		\begin{equation}
			\mathcal{P}^2 = \mathcal{P} = \mathcal{P}^*.
		\end{equation}
		\item\label{it:spectral} Let the spectral decomposition of $ H_0 $ have the form  $ H_0 = \sum_{\varepsilon}\varepsilon \Pi_{\varepsilon} $, where $ \varepsilon $ are (distinct) eigenvalues of $ H_0 $ and $ \Pi_{\varepsilon}  $ are orthogonal projectors $ \Pi_{\varepsilon}  \Pi_{\varepsilon'} = \delta_{\varepsilon \varepsilon'}  \Pi_{\varepsilon} $, $  \Pi_{\varepsilon} =  \Pi_{\varepsilon}^{\dagger} $. Then
		\begin{equation}\label{eq:PProjRep}
			\mathcal{P} X = \sum_{\varepsilon} \Pi_{\varepsilon}  X \Pi_{\varepsilon}
		\end{equation}
		for any matrix $ X $.
	\end{enumerate}
	For the case of 1-dimensional projectors $ \Pi_{\varepsilon} $, \eref{eq:PProjRep} is sometimes called the dephasing operation \cite{streltsovColloquiumQuantumCoherence2017}. In the general case it is usually called pinching \cite[p. 16]{tomamichel2015quantum}. It can also be understood as a special case of twirling \cite{bennett1996mixed} (with one-parameter group).
	
	Effective Hamiltonian $  \tilde{H} $ can be calculated by cumulant-type expansion. Namely, we have the following proposition (see \ref{app:proofOfCumulant} for the proof).
	
	\begin{proposition}\label{prop:effectiveHamExp}
		The perturbative expansion of $ \tilde{H} $ has the form
		\begin{equation}\label{eq:effectiveHamExp}
			\tilde{H} = H_0 - \beta^{-1}  \sum_{n=1}  \lambda^n \sum_{k_0 + \cdots + k_m = n} \frac{(-1)^m}{m + 1} \mathcal{M}_{k_0}(\beta) \mathcal{M}_{k_1}(\beta) \cdots \mathcal{M}_{k_m}(\beta),
		\end{equation}
		where
		\begin{equation}\label{eq:moments}
			\mathcal{M}_{k}(\beta) = (-1)^k \int_{0}^{\beta} d \beta_1  \ldots  \int_{0}^{\beta_{k-1}} d \beta_k  \mathcal{P} H_{I}(\beta_1) \ldots H_{I}(\beta_k)
		\end{equation}
		and
		\begin{equation}\label{eq:intHamIntPic}
			H_{I}(\beta) \equiv e^{\beta H_0} H_I e^{-\beta H_0}.
		\end{equation}
	\end{proposition}
	
	In particular, the first terms of the expansion have the form
	\begin{equation}\label{eq:firstTerms}
		\tilde{H} = H_0 - \beta^{-1}  \lambda \mathcal{M}_{1}(\beta)  - \beta^{-1} \lambda^2 \left(\mathcal{M}_{2}(\beta) - \frac12 (\mathcal{M}_{1}(\beta))^2\right)  + O(\lambda^3).
	\end{equation}
	
	To make this expansion more explicit, let us represent the interaction Hamiltonian in the eigenbasis of the superoperator $ [H_0, \; \cdot \;] $ in the same way as it is usually done for Markov master equation derivation \cite[subsection~3.3.1]{breuerTheoryOpenQuantum2007}
	\begin{equation}\label{eq:HIEigenOperExp}
		H_I = \sum_{\omega} D_{\omega},
	\end{equation}
	where sum is taken over the Bohr frequencies and
	\begin{equation}\label{eq:HIEigenOper}
		[H_0, D_{\omega}] = - \omega D_{\omega}.
	\end{equation}
	Moreover, as $ H_I $ is Hermitian, then $ D_{-\omega} = D_{\omega}^{\dagger} $. Hence, we have the following explicit expressions for $ \mathcal{M}_{k}(\beta) $.
	\begin{proposition}\label{prop:EigenOperExp}
		If \eref{eq:HIEigenOperExp}-\eref{eq:HIEigenOper} are held, then 
		\begin{equation}\label{en:EigenOperExp}
			\fl
			\mathcal{M}_{k}(\beta) = (-1)^k \sum_{\omega_1, \ldots, \omega_{k-1}} g_{k}(\beta; \omega_1, \cdots,  \omega_{k-1})  D_{\omega_1} \cdots D_{\omega_{k-1}} D_{-\omega_1 - \ldots - \omega_{k-1}},
		\end{equation}
		where
		\begin{eqnarray}
			&g_{k}(\beta; \omega_1, \ldots,  \omega_{k-1})  =\frac{1}{\prod _{k=1}^{n-1} \sum _{j=1}^k \omega_j} \left(\beta - \sum _{k=1}^{n-1} \frac{1}{\sum _{j=1}^k \omega _j} \right) \nonumber\\ 
			&- \sum_{p=1}^{n-1} \frac{(-1)^p }{\left(\prod_{m=2}^p \sum_{i=m}^p \omega _i\right) \left(\sum_{r=1}^p \omega_r\right)^2 
				\left(\prod _{k=p+1}^{n-1} \sum _{j=p+1}^k
				\omega _j\right)} e^{-\beta \sum_{i=1}^p \omega_i}.
		\end{eqnarray}
		For zero denominators it should be understood as a limit.
	\end{proposition}
	The proof can be found in \ref{app:proofOfExplicit}. The first terms of expansion \eref{eq:firstTerms} take the form (see  \ref{app:proofOfExplicit})
	\begin{equation}\label{eq:expanToSec}
		\tilde{H} = H_0 + \lambda  D_0 - \lambda^2 \sum_{\omega \neq 0} \frac{\beta  \omega +e^{-\beta \omega }-1}{\beta \omega ^2} D_{\omega} D_{\omega}^{\dagger} + O(\lambda^3)
	\end{equation}
	Thus, the first two terms are temperature-independent and recover the Hamiltonian in the rotating wave approximation (similarly to effective Hamiltonians for dynamical evolution \cite{trubilko2020hierarchy, basharovEffectiveHamiltonianNecessary2021})
	\begin{equation}
		H_{\rm RWA} =  H_0 + \lambda  D_0.
	\end{equation}
	
	On the other hand the next term of expansion \eref{eq:expanToSec} is the first temperature-dependent correction to RWA Hamiltonian. This term is always non-positive definite
	\begin{equation}
		\tilde{H}^{(2)} \equiv - \sum_{\omega \neq 0} \frac{\beta  \omega +e^{-\beta \omega }-1}{\beta \omega ^2}  D_{\omega} D_{\omega}^{\dagger}  \leq 0
	\end{equation}
	due to the fact that it has the form
	\begin{equation}\label{eq:secOrderHeffTerm}
		\tilde{H}^{(2)} = -\frac{\beta}{2} \sum_{\omega \neq 0} f(\beta \omega)  D_{\omega} D_{\omega}^{\dagger},
	\end{equation}
	where $ \langle \psi |D_{\omega} D_{\omega}^{\dagger} | \psi \rangle =|| D_{\omega}^{\dagger} | \psi \rangle ||^2 \geq 0 $ for arbitrary $ | \psi \rangle $,
	\begin{equation}\label{eq:fDef}
		f(x) \equiv 2 \frac{x +e^{-x} -1}{x^2}
	\end{equation}
	is a positive function $ f(x) > 0 $ for all real $ x $ and $ \beta $ is assumed to be positive as we consider the positive temperature (but if one considers a negative temperature, which is possible for finite-dimensional systems, then $ \tilde{H}^{(2)}  $ becomes non-negative). Moreover, $ \tilde{H}^{(2)}  $ is a monotone function of temperature, because
	\begin{equation}\label{en:monotoneSecOrder}
		\frac{\partial}{\partial \beta}\tilde{H}^{(2)} = -\frac{1}{2} \sum_{\omega \neq 0} f_1(\beta \omega)  D_{\omega} D_{\omega}^{\dagger} \leq 0,
	\end{equation}
	where
	\begin{equation}
		f_1(x) \equiv 2 \frac{1- e^{-x}(1+x)}{x^2}
	\end{equation}
	is also a positive function for all real  $ x $.  In the next section we will see that if one averages this result with respect to the effective Gibbs state, then this result becomes closely related to general thermodynamic properties which are valid in all the orders of perturbation theory.
	
	Let us also remark that $ \lim\limits_{x \rightarrow + 0} f(x) = 1$, so for the low temperature limit, i.e. when $ \beta \omega \gg 1 $ for all non-zero Bohr frequencies,   \eref{en:monotoneSecOrder} takes the form
	\begin{equation}
		\tilde{H}^{(2)} \approx -\frac{\beta}{2} \sum_{\omega \neq 0}  D_{\omega} D_{\omega}^{\dagger},
	\end{equation}
	i.e. the second order correction in $ \lambda $ is linear in $ \beta $.
	
	In recent literature there is also rising interest in the ultrastrong coupling limit. Let us remark that $ \tilde{H}^{(2)} $ is also the leading order difference between effective Hamiltonians for steady states for the ultrastrong coupling limit conjectured in \cite{goyalSteadyStateThermodynamics2019} and the one obtained in \cite{cresserWeakUltrastrongCoupling2021a}, if one takes the interaction Hamiltonian as a free Hamiltonian in our notation and vice versa. The perturbative corrections for such steady states are discussed in \cite{latuneSteadyStateUltrastrong2021}. 
	
	\section{Effective Hamiltonian as analog of mean force Hamiltonian}\label{sec:tempDepEff}
	
	The free energy $ F $ can be defined by the partition function $ Z $ as
	\begin{equation}
		F = - \beta^{-1} \ln Z,
	\end{equation}
	where, as it was mentioned before, $ Z $ could be defined by the same formula $ Z = \Tr e^{- \beta H} = \Tr e^{- \beta \tilde{H}} $ both by exact Hamiltonian $ H $ and by effective Hamiltonian $ \tilde{H} $. If one calculates the entropy and the internal energy by equilibrium thermodynamics formulae
	\begin{equation}
		S = \beta^2 \frac{\partial F}{\partial \beta}, \qquad U = \frac{\partial (\beta F)}{\partial \beta},
	\end{equation}
	then it also obviously does not matter if we use the exact or effective Hamiltonian. For initial, temperature-independent Hamiltonian they also could be calculated as:
	\begin{equation}\label{eq:exactSU}
		S = - \Tr \rho_{\beta} \ln  \rho_{\beta}, \qquad U = \Tr H \rho_{\beta}.
	\end{equation}
	
	But for the effective Hamiltonian the similar formulae need additional corrections due to its dependence on temperature. Namely, 
	\begin{equation}
		S = \tilde{S} - \Delta S, \qquad 	U = \tilde{U}  -  \Delta U,
	\end{equation}
	where $ \tilde{S} $ and $ \tilde{U} $ are defined by the formulae similar to \eref{eq:exactSU}
	\begin{equation}\label{eq:observEntropyEnergy}
		\tilde{S} = - \Tr \tilde{\rho}_{\beta} \ln  \tilde{\rho}_{\beta}, \qquad \tilde{U} = \Tr \tilde{H} \tilde{\rho}_{\beta}
	\end{equation}
	And the corrections have exactly the same form as for the mean force Hamiltonian (see, e.g. \cite{rivasStrongCouplingThermodynamics2020},   (11)-(12))
	\begin{equation}\label{eq:deltaEntropyEnergy}
		\Delta S = - \beta^2 \langle \partial_{\beta} \tilde{H} \rangle_{\sim}, \qquad \Delta U = - \beta \langle \partial_{\beta} \tilde{H} \rangle_{\sim} = \beta^{-1} \Delta S.
	\end{equation}
	Here $ \langle \; \cdot \; \rangle_{\sim} $ denotes the average with respect to the effective Gibbs state, i.e. $ \langle \; \cdot \; \rangle_{\sim} \equiv \Tr (\; \cdot \; \tilde{\rho}_{\beta} )   $. The derivation of these formulae is exactly the same as for analogous formulae for the mean force  Hamiltonian (see  \cite{seifertFirstSecondLaw2016}, \cite[chapter 22]{binderThermodynamicsQuantumRegime2018}), because it is  valid for an arbitrary temperature-dependent Hamiltonian and is based only on the Feynman-Wilcox formula \cite{feynmanOperatorCalculusHaving1951, wilcoxExponentialOperatorsParameter2004, chebotarevOperatorvaluedODEsFeynman2012} 
	\begin{equation}
		\frac{d}{d \beta} e^{-\beta \tilde{H}} = -\int_0^t ds e^{-(1-s)\beta \tilde{H}} \left(\frac{d}{d \beta} (\beta \tilde{H})\right)  e^{-s\beta \tilde{H}}.
	\end{equation}
	
	Due to the fact that $ \mathcal{P} $  is a completely positive trace preserving and unital map ($ \mathcal{P} I = I $) the entropy is monotone \cite[p. 136]{holevoQuantumSystemsChannels2012} under its action, i.e.  $ \tilde{S} \geq S $. So $ \Delta S \geq 0 $ and $ \Delta U = \beta^{-1} \Delta S \geq 0 $. $ \tilde{S}  $ and $ \tilde{U} $ could be interpreted as entropy and as energy which are accessible to our observations. Our observable entropy is $  \tilde{S} $, but due to our restricted observational capabilities we have the information loss quantified by $ \Delta S $. This information loss comes with energy loss quantified by $ \Delta U $ and is hidden from our observations.
	
	For second order expansion in $ \lambda $ we have 
	\begin{equation}\label{eq:deltaEntropySecOrd}
		\Delta S =  -\lambda^2 \beta^2 \langle \partial_{\beta} \tilde{H}^{(2)} \rangle_{\sim}  + O(\lambda^3)  =  -\lambda^2 \beta^2 \langle \partial_{\beta} \tilde{H}^{(2)} \rangle_{0}  + O(\lambda^3),
	\end{equation}
	where $  \langle  \; \cdot \; \rangle_{0}  $ is the average with respect to the Gibbs state for the free Hamiltonian. So the non-negativity of $ \Delta S $ in the second order of perturbation theory agrees with \eref{en:monotoneSecOrder}. Moreover, it could be calculated (see  \ref{app:simplCalc}) by the following formula
	\begin{equation}\label{eq:deltaEntropySecOrdSimp}
		\Delta S  =  -\lambda^2 \beta \langle \tilde{H}^{(2)} \rangle_{0}  + O(\lambda^3) = \sum_{\omega > 0} \frac{1 - e^{- \beta \omega }}{\beta \omega} \langle D_{\omega} D_{\omega}^{\dagger} \rangle_0 + O(\lambda^3),
	\end{equation} 
	where sum is taken only over the positive Bohr frequencies.

	The analogy with ~(22.6) of \cite[chapter 22]{binderThermodynamicsQuantumRegime2018} also suggests the following definition of non-equilibrium free energy in a given state $ \rho $
	\begin{equation}\label{eq:nonEqObsFreeEn}
		\tilde{F}_{\rho} \equiv \langle \tilde{H} \rangle_{\mathcal{P}} + \beta^{-1} \langle \ln \mathcal{P} \rho \rangle_{\mathcal{P}} = F + \beta^{-1} S(\mathcal{P}\rho||\tilde{\rho}_{\beta}),
	\end{equation} 
	where $  \langle \; \cdot \; \rangle_{\mathcal{P}} \equiv \Tr(\mathcal{P}\rho \; \cdot \;)  $ and $ S(\rho || \sigma) $  is relative entropy \cite[chapter~7.1]{holevoQuantumSystemsChannels2012}. The only difference from ~(22.6) of \cite[chapter 22]{binderThermodynamicsQuantumRegime2018} consists in the fact that we use averaged state $ \mathcal{P} \rho $ instead of $ \rho $, which is natural in our setup.
	
	The exact free energy is defined as
	\begin{equation}\label{eq:nonEqExactFreeEn}
		F_{\rho} \equiv \langle H \rangle + \beta^{-1} \langle \ln \rho \rangle = F + \beta^{-1} S(\rho||\rho_{\beta}),
	\end{equation} 
	where $  \langle \; \cdot \; \rangle \equiv \Tr(\rho \; \cdot \;)  $, which leads to
	\begin{equation}
		F_{\rho} = \tilde{F}_{\rho} + \Delta F_{\rho},
	\end{equation}
	where similarly to \eref{eq:deltaEntropyEnergy} $ \Delta F_{\rho} $ has a definite sign, namely
	\begin{equation}
		\Delta F_{\rho} =\beta^{-1}  (S(\rho||\rho_{\beta}) -  S(\mathcal{P}\rho|| \mathcal{P}\rho_{\beta})) \geq 0
	\end{equation}
	due to monotonicity of the relative entropy under the completely positive map $ \mathcal{P} $ \cite[theorem~7.6]{holevoQuantumSystemsChannels2012}. Similarly to $ \tilde{S} $ and $ \tilde{U} $, $ \tilde{F}_{\rho}  $ can be interpreted as observable free energy and $ \Delta F_{\rho} $ as free energy hidden from our observations. As $ \Delta F_{\rho} \geq 0$, we are always further from equilibrium than we think based on our restricted measurement possibilities. For example, if our exact non-equilibrium state is $ \tilde{\rho}_{\beta} $, then it is impossible to distinguish it from $ \rho_{\beta} $. So its observable free energy coincides with the equilibrium one   
	\begin{equation}
		\tilde{F}_{\tilde{\rho}_{\beta}} = F +  \beta^{-1} S(\tilde{\rho}_{\beta}||\tilde{\rho}_{\beta})  = F,
	\end{equation} 
	but $ \Delta F_{\tilde{\rho}_{\beta}}  $ is positive as in the general case. Namely, by  \eref{eq:nonEqObsFreeEn}-\eref{eq:nonEqExactFreeEn} we have
	\begin{equation}
		\Delta F_{\tilde{\rho}_{\beta}} =  \langle H \rangle_{\sim} - \langle \tilde{H} \rangle_{\sim}.
	\end{equation}
	As $ \langle H \rangle_{\sim} = \Tr H \mathcal{P} \rho = \Tr H \mathcal{P} \mathcal{P} \rho_{\beta} = \Tr \mathcal{P}( H ) \mathcal{P} \rho_{\beta}  = \Tr H_{\rm RWA} \rho_{\beta} = \langle H_{\rm RWA} \rangle_{\sim} $, then
	\begin{equation}
		\Delta F_{\tilde{\rho}_{\beta}} =  \langle H_{\rm RWA} - \tilde{H} \rangle_{\sim}.
	\end{equation}
	This formula is useful for asymptotic expansion of $ \Delta F_{\tilde{\rho}_{\beta}}  $ as the first two terms of the expansion of $ \tilde{H} $ cancel $ H_{\rm RWA}  $  and the first non-trivial contribution is of order of $ \lambda^2 $ as in \eref{eq:deltaEntropySecOrd}. Namely, we have
	\begin{equation}
		\Delta F_{\tilde{\rho}_{\beta}} = - \lambda^2 \langle H^{(2)} \rangle_{\sim}  + O(\lambda^3) = - \lambda^2 \langle H^{(2)} \rangle_{0}  + O(\lambda^3).
	\end{equation}
	Moreover, it is possible to show (see  \ref{app:simplCalc}) that $ \langle \partial_{\beta} H^{(2)} \rangle_{0} = \beta^{-1} \langle H^{(2)} \rangle_{0}$, so
	\begin{equation}
		\Delta U = \Delta F_{\tilde{\rho}_{\beta}} + O(\lambda^3), \qquad \Delta S = \beta \Delta F_{\tilde{\rho}_{\beta}} + O(\lambda^3).
	\end{equation}

	The analogy with the mean force Hamiltonian can be made more  explicit if one notes that the mean force Hamiltonian is closely related to the projector $ \mathcal{P}' = \Tr_B(\; \cdot \;) \otimes \rho_B$ which is usually used for derivation of Markovian master equations and their perturbative corrections \cite[subsection 9.1.1]{breuerTheoryOpenQuantum2007}.
	\begin{equation}
		\mathcal{P}' \frac{e^{- \beta H}}{Z}= \frac{1}{Z} \Tr_B e^{- \beta H} \otimes \frac{1}{Z_B}e^{ -\beta H_B} = \frac{1}{Z_{\rm mf}} e^{- \beta H_{\rm mf}} \otimes \frac{1}{Z_B}e^{ -\beta H_B},
	\end{equation}
	where $ Z_{\rm mf} = Z/Z_B $ \cite{talknerColloquiumStatisticalMechanics2020a}. Thus, a stricter analog of our effective Hamiltonian should be $ H_{\rm mf} + H_B$ with partition function $ Z $. But it seems that for operational meaning of the mean force Hamiltonian the information about $ H_B $ is also important, which makes this analog more natural. Nevertheless,  importance of information about $ H_B $ (not $ H_{\rm mf} $ only) is still discussible \cite{talknerCommentMeasurabilityNonequilibrium2020,strasbergMeasurabilityNonequilibriumThermodynamics2020}.
	
	From the mathematical point of view both of these projectors are so-called conditional expectations \cite{nakamuraRemarkExpectationsOperator1960, umegakiConditionalExpectationOperator1962, accardiConditionalExpectationsNeumann1982, dominyDualityConditionalExpectations2017}. They are correspondent to different choices of observable degrees of freedom. This suggests that the mean force Hamiltonian theory could be generalized to arbitrary conditional expectations and for specific conditional expectation $  \mathcal{P} $ it is done in this work. Thus, it is possible to say that the effective Gibbs state with such generalized projectors define different effective quantum equilibrium thermodynamics.

		\section{Mean force Hamiltonian for effective Gibbs state}\label{sec:meanForce}
	
	Let us now consider a compound system, consisting of two subsystems $ A $ and $ B $. Let us consider subsystem $ B $ as ''reservoir''.  Let us assume that $ H_0 = H_A \otimes I + I \otimes H_B $.  Then it is possible to define a mean for the Hamiltonian $ \tilde{H}_{\rm mf} $ for the effective Gibbs state by the following formula
	\begin{equation}
		\tilde{\rho}_{\rm mf} \equiv \Tr_B \mathcal{P}\rho_{\beta} = \frac{1}{\tilde{Z}_{\rm mf}} e^{- \beta \tilde{H}_{\rm mf}},
	\end{equation}
	where $  \tilde{Z}_{\rm mf} = \tilde{Z}/Z_B $, $ Z_B \equiv \Tr_B e^{-\beta H_B} $. Then similarly to proposition \ref{prop:effectiveHamExp} it is possible to obtain the perturbative expansion in $ \lambda $ for $ \tilde{H}_{\rm mf} $ (see  \ref{sec:meanForceEffGibbs}).
	
	\begin{proposition}\label{prop:meanForce}
		The perturbative expansion of $ \tilde{H}_{\rm mf} $ in $ \lambda $ has the form
		\begin{equation}
			\fl
			\tilde{H}_{\rm mf} = H_A - \beta^{-1}  \sum_{n=1}^{\infty}  \lambda^n \sum_{k_0 + \cdots + k_m = n} \frac{(-1)^m}{m + 1} \langle \mathcal{M}_{k_0}(\beta)  \rangle_{B} \langle \mathcal{M}_{k_1}(\beta) \rangle_{B} \cdots \langle \mathcal{M}_{k_m}(\beta) \rangle_{B},
		\end{equation}
		where $ \langle \; \cdot \; \rangle_B  \equiv \Tr_B( \; \cdot \;  Z_B^{-1} e^{-\beta H_B})$.
	\end{proposition}
	Here  $  \mathcal{M}_{k}(\beta) $ can be also calculated by proposition \ref{prop:EigenOperExp}. The first terms of the expansion for $ 	\tilde{H}_{\rm mf} $ have the form
	\begin{equation}\label{eq:meanForcSecOrder}
		\fl
		\tilde{H}_{\rm mf} = H_A + \lambda \langle  D_0 \rangle_{B} - \lambda^2 \frac{\beta}{2}\left(\sum_{\omega \neq 0} f(\beta \omega)\langle  D_{\omega} D_{\omega}^{\dagger} \rangle_{B} + \langle D_0^2 \rangle_B - \langle D_0 \rangle_B^2\right) + O(\lambda^3).
	\end{equation}
	
	This formula can be made even more explicit if one considers the decomposition of  $ D_{\omega} $ into sum of eigenoperators of $ [H_A, \; \cdot \;] $ and $ [H_B, \; \cdot \;] $, i.e. similarly to \eref{eq:HIEigenOper} introducing $ A_{\omega} $ and $ B_{\omega} $ such that 
	\begin{equation}\label{eq:subEigenProj}
		[H_A, A_{\omega_1}] = - \omega_1 A_{\omega_1}, \qquad [H_B, B_{\omega_2}] = - \omega_2 B_{\omega_2},
	\end{equation}
	where $ \omega_1 $ and $ \omega_2 $ run over all possible Bohr frequencies of the Hamiltonians $ H_A $ and $ H_B $, respectively. Then expansion \eref{eq:meanForcSecOrder} takes the form (see \ref{app:meanForcSecOrderExplicit})
	\begin{eqnarray}
		\fl
		\tilde{H}_{\rm mf} = H_S + \lambda   \langle  B_{0} \rangle_{B} A_{0}  
		- \lambda^2 \frac{\beta}{2 } \biggl(\sum_{\omega_1 \neq 0} \left( \sum_{\omega} f(\beta \omega )   \langle  B_{\omega_1 + \omega} B_{\omega_1 + \omega}^{\dagger} \rangle_B \right)   A_{ \omega_1 }^{\dagger}  A_{\omega_1}	\nonumber \\
		+    \left( \sum_{\omega}  f(\beta \omega ) \langle  B_{\omega} B_{ \omega}^{\dagger} \rangle_B  - \langle  B_{0} \rangle_B^2\right) A_{0}^2  \biggr) + O(\lambda^3),
		\label{eq:meanForcSecOrderExplicit}
	\end{eqnarray}
	where it is assumed that $ f(0)=1 $.

	\section{Examples}\label{sec:examples}
	
	In this section we consider several examples and the notations are chosen in such a way as to emphasize the similarity between them. We use these examples to illustrate our formulae, but let us remark that at least for the first and second model it is possible to calculate effective Hamiltonian exactly without perturbation theory, however it is not the aim of our work. For all these examples we consider two cases: the off-resonance and the resonance one. In this section only the results are presented, all the calculations are given separately in  \ref{app:calcForExamp}.
	
	\subsection{Two interacting two-level systems}
	
	Let us consider the two interacting two-level systems \cite{lisenfeldObservationDirectlyInteracting2015, trushechkinPerturbativeTreatmentIntersite2016}
	$ a $ and $ b $
	\begin{equation}\label{eq:twoTwoLevel}
		H = \omega_a \sigma_a^+\sigma_a^- + \omega_b \sigma_b^+\sigma_b^- + \lambda (\sigma_a^- + \sigma_a^+)  (g^* \sigma_b^-  + g \sigma_b^+) ,
	\end{equation}
	where $ \omega_a  > 0 $, $ \omega_b  > 0 $ and $  \sigma_i^{\pm} $ are usual ladder operators for two-level  systems $ i =a,b $.
	
	1) Off-resonance case $ \omega_a  \neq \omega_b $.
	\begin{eqnarray}
		\fl
		\tilde{H}_{\rm off-res} = \omega_a  n_a + \omega_b  n_b - \lambda^2 \frac{\beta}{2} |g|^2 (f(\beta(\omega_a - \omega_b)) (1-n_a) n_b \nonumber\\
		+ f(\beta(\omega_a + \omega_b)) (1-n_a) (1 - n_b) 
		+ f(\beta(\omega_b -\omega_a))  n_a(1-n_b) \nonumber\\
		+ f(\beta(-\omega_a  - \omega_b)) n_a n_b) + O(\lambda^3), \label{eq:twoTwoLevelEff}
	\end{eqnarray}
	where $ n_i \equiv \sigma_i^+\sigma_i^- $ are number operators for $ i =a,b $. In the leading order the information loss has the form
	\begin{equation}\label{eq:offResEnt}
		\Delta S_{\rm off-res} = \lambda^2 \beta |g|^2 \frac{\omega_a  \tanh \frac{\beta \omega_a }{2} - \omega_b \tanh \frac{\beta  \omega_b}{2}}{\omega_a^2 - \omega_b^2} + O(\lambda^3).
	\end{equation}
	
	2) Resonance case $ \omega_b = \omega_a +  \lambda \delta \omega $.
	\begin{eqnarray}
		\fl
		\tilde{H}_{\rm res} = \omega_a n_a + \omega_b n_b + \lambda (g\sigma_a^- \sigma_b^+ +  g^*\sigma_a^+ \sigma_b^-)  \nonumber\\
		-\lambda^2 \frac{\beta}{2} |g|^2 (f(2 \beta \omega_a) (1-n_a) (1 - n_b) + f(-2 \beta \omega_a) n_a n_b) + O(\lambda^3).\label{eq:twoTwoLevelEffRes}
	\end{eqnarray}
	In the leading order the information loss has the form
	\begin{equation}\label{eq:resEnt}
		\Delta S_{\rm res} = \lambda^2 \beta |g|^2 \frac{\tanh \frac{\omega_a \beta}{2}}{2 \omega_a}  + O(\lambda^3).
	\end{equation}
	Let us remark that it does not coincide with the off-resonance case with $ \omega_b \rightarrow \omega_a $. Namely, we have
	\begin{equation}\label{eq:offResEntAndResEnt}
		\Delta S_{\rm off-res}|_{\omega_b \rightarrow \omega_a} = \Delta S_{\rm res} + \lambda^2 \left(\frac{\beta |g|}{2 \cosh \frac{\beta \omega_a}{2}}\right)^2 + O(\lambda^3).
	\end{equation}
	Thus, off-resonance averaging leads to larger information loss even in the ''resonance'' limit than resonance averaging.

	\begin{figure}[t]
		\includegraphics{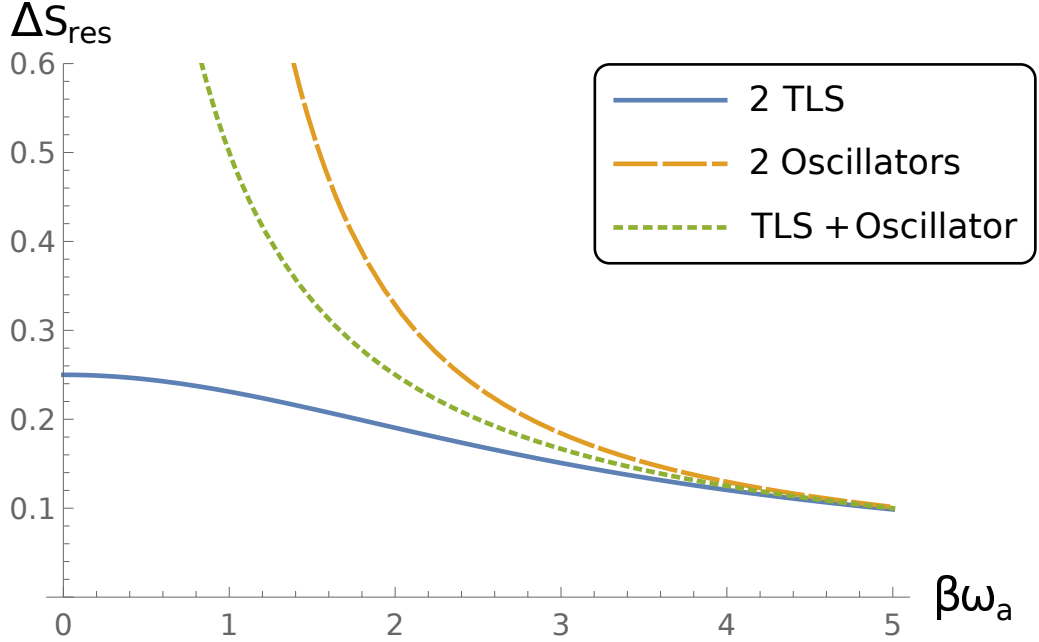}
		\caption{The information loss for resonance case and $ \beta |g| = 1 $ for two two-level systems (solid line), two oscillators (dashed line) and two-level system interaction with oscillator (dotted line).} 
		\label{fig}
	\end{figure}
	
	\subsection{Two interacting harmonic oscillators}
	
	Let us consider the two interacting harmonic oscillators
	\begin{equation}
		H = \omega_a a^{\dagger} a + \omega_b b^{\dagger} b + \lambda (a + a^{\dagger})  (g^* b + g b^{\dagger}) ,
	\end{equation}
	where $ \omega_a  > 0 $, $ \omega_b  > 0 $ and $ a, a^{\dagger} $ and $ b, b^{\dagger} $ are oscillator (bosonic) ladder operators. Averaging with respect to fast oscillations needed for so-called quasi-stationary states was recently discussed in \cite{lopez-saldivarDifferentialParametricFormalism2020}
	
	1) Off-resonance case $ \omega_a  \neq \omega_b $.
	\begin{eqnarray}
		\fl
		\tilde{H}_{\rm off-res} = \omega_a  n_a + \omega_b  n_b - \lambda^2 \frac{\beta}{2} |g|^2 (f(\beta(\omega_a + \omega_b)) (1+n_a) n_b \nonumber \\
		 + f(\beta(\omega_a + \omega_b)) (1+n_a) (1 + n_b)
		+ f(\beta(\omega_b -\omega_a))  n_a(1+n_b)  \nonumber\\
		 + f(\beta(-\omega_a  - \omega_b)) n_a n_b) + O(\lambda^3), \label{eq:twoOscEff}
	\end{eqnarray}
	where $ n_a \equiv a^{\dagger} a $, $ n_b \equiv b^{\dagger} b $. In the leading order the information loss has the form
	\begin{equation}\label{eq:offResEntOsc}
		\Delta S_{\rm off-res} = \lambda^2 \beta |g|^2 \frac{\omega_a  \coth \frac{\beta \omega_b }{2} - \omega_b \coth \frac{\beta  \omega_a}{2}}{\omega_a^2 - \omega_b^2} + O(\lambda^3).
	\end{equation}
	
	2) Resonance case $ \omega_b = \omega_a +  \lambda \delta \omega $.
	\begin{eqnarray}
		\fl
		\tilde{H}_{\rm res} = \omega_a n_a + \omega_b n_b + \lambda (g a b^{\dagger} +  g^* a^{\dagger} b) \nonumber \\
		-\lambda^2 \frac{\beta}{2} |g|^2 (f(2 \beta \omega_a) (1 + n_a) (1 + n_b) + f(-2 \beta \omega_a) n_a n_b) + O(\lambda^3).  \label{eq:twoOscEffRes}
	\end{eqnarray} 
	In the leading order the information loss has the form
	\begin{equation}\label{eq:resEntOsc}
		\Delta S_{\rm res} = \lambda^2 \beta |g|^2 \frac{\coth \frac{\omega_a \beta}{2}}{2 \omega_a}  + O(\lambda^3).
	\end{equation}
	Interestingly, this quantity asymptotically coincides with \eref{eq:resEnt} for $ \omega_a \beta \gg 1 $ (see figure~\ref{fig}). Similarly to \eref{eq:offResEntAndResEnt} we have
	\begin{equation}
		\Delta S_{\rm off-res}|_{\omega_b \rightarrow \omega_a} = \Delta S_{\rm res} + \lambda^2 \left(\frac{\beta |g|}{2 \sinh \frac{\beta \omega_a}{2}}\right)^2 + O(\lambda^3).
	\end{equation}

	\subsection{Two-level system interacting with harmonic oscillator}
	Let us consider a two level system interacting with a harmonic oscillator
	\begin{equation}
		H = \omega_a \sigma^+ \sigma^- + \omega_b b^{\dagger} b + \lambda (\sigma^+ + \sigma^-)  (g^* b + g b^{\dagger}) ,
	\end{equation}
	where $ \omega_a  > 0 $, $ \omega_b  > 0 $ and $ \sigma^+, \sigma^-  $ and $ b, b^{\dagger} $ are two-level and bosonic ladder operators, respectively.
	
	1) Off-resonance case $ \omega_a  \neq \omega_b $.
	\begin{eqnarray}
		\fl
		\tilde{H}_{\rm off-res} = \omega_a  n_a + \omega_b  n_b - \lambda^2 \frac{\beta}{2} |g|^2 (f(\beta(\omega_a + \omega_b)) (1-n_a) n_b  \nonumber\\
		+ f(\beta(\omega_a + \omega_b)) (1-n_a) (1 + n_b)
		+ f(\beta(\omega_b -\omega_a))  n_a(1+n_b)  \nonumber\\
		+ f(\beta(-\omega_a  - \omega_b)) n_a n_b) + O(\lambda^3), \label{eq:twoLevelOscEff}
	\end{eqnarray}
	where $ n_a \equiv  \sigma^+ \sigma_- $, $ n_b \equiv b^{\dagger} b $. In the leading order the information loss has the form
	\begin{equation}\label{eq:offResEntTLSOsc}
		\Delta S_{\rm off-res} = \lambda^2 \beta |g|^2 \frac{\omega_a  \tanh \frac{\beta  \omega_a}{2}  \coth \frac{\beta \omega_b }{2}- \omega_b }{\omega_a^2 - \omega_b^2} + O(\lambda^3).
	\end{equation}
	
	2) Resonance case $ \omega_b = \omega_a + \lambda \delta \omega $.
	\begin{eqnarray}
		\fl
		\tilde{H}_{\rm res} = \omega_a n_a + \omega_b n_b + \lambda (g \sigma^- b^{\dagger} +  g^* \sigma^+ b)  \nonumber \\
		-\lambda^2 \frac{\beta}{2} |g|^2 (f(2 \beta \omega_a) (1 - n_a) (1 + n_b) + f(-2 \beta \omega_a) n_a n_b) + O(\lambda^3).
	\end{eqnarray} 
	In the leading order the information loss has the form
	\begin{equation}\label{eq:resEntTLSOsc}
		\Delta S_{\rm res} = \lambda^2 \frac{\beta |g|^2}{2 \omega_a} + O(\lambda^3).
	\end{equation}
	This also asymptotically coincides with \eref{eq:resEnt} for $ \omega_a \beta \gg 1 $ (see figure \ref{fig}). Similarly to \eref{eq:offResEntAndResEnt} we have
	\begin{equation}
		\Delta S_{\rm off-res}|_{\omega_b \rightarrow \omega_a} = \Delta S_{\rm res} + \lambda^2 \frac{\beta |g|^2}{2 \sinh \beta \omega_a} + O(\lambda^3).
	\end{equation}

	\section{Conclusions}
	We have developed a systematic perturbative calculation of the effective Hamiltonian which defines the effective Gibbs state for the averaged observables. We have shown that the first two terms of the perturbative expansion of such an effective Hamiltonian coincide with RWA Hamiltonian and the second order term of the expansion  is the first non-trivial temperature-dependent term. It defines the leading order of the  information loss  due to restricted observation capabilities in this setup and the leading order of the energy, which is not observable in our setup due to the same reason. We have shown the analogy between our setup and the mean force Hamiltonian. To deepen this analogy we have also obtained the perturbative expansion for the mean force Hamiltonian for the effective Gibbs state. At the end we have considered several examples, which illustrate the preceding material.
	
	We think that the analogy between the mean force Hamiltonian and our effective Hamiltonians suggests the possibility to generalize our approach to form effective equilibrium quantum thermodynamics. 
	
	\ack
	
	The author thanks A. S. Trushechkin for the fruitful discussion of the problems considered in the work. This work was funded by Russian Federation represented by the Ministry of Science and Higher Education (grant number 075-15-2020-788).
	
	\appendix
	
	\section{Properties of averaging projector}
	\label{app:avProj}
	
	Trace preservation of $ \mathcal{P} $ follows from 
	\begin{equation}
		\fl
		\Tr \mathcal{P}  X  = \lim\limits_{T \rightarrow + \infty} \frac{1}{T}\int_0^T \Tr e^{i H_0 t} X e^{-i H_0 t} dt =  \lim\limits_{T \rightarrow + \infty} \frac{1}{T}\int_0^T dt \Tr X  = \Tr X,
	\end{equation}
	Then let us prove property \eref{it:spectral} first. For $ H_0 = \sum_{\varepsilon}\varepsilon \Pi_{\varepsilon} $  we have
	\begin{eqnarray}
		\fl
		\mathcal{P}  X  = \lim\limits_{T \rightarrow + \infty} \frac{1}{T}\int_0^T dt e^{i H_0 t}  X e^{-i H_0 t}  \nonumber \\
		= \lim\limits_{T \rightarrow + \infty} \frac{1}{T}\int_0^T  dt  \sum_{\varepsilon, \varepsilon'} e^{i (\varepsilon - \varepsilon') t} \Pi_{\varepsilon}  X \Pi_{\varepsilon'} =  \sum_{\varepsilon} \Pi_{\varepsilon}  X \Pi_{\varepsilon}
	\end{eqnarray}
	since
	\begin{equation}
		\lim\limits_{T \rightarrow + \infty} \frac{1}{T}\int_0^T  dt e^{i (\varepsilon - \varepsilon') t} =  \delta_{\varepsilon \varepsilon'} .
	\end{equation}
	As $ \Pi_{\varepsilon} = \Pi_{\varepsilon}^{\dagger} $, then they define Kraus representation \cite[p. 110]{holevoQuantumSystemsChannels2012} of $ \mathcal{P}  $, which proves property \eref{it:complPos}. Calculating
	\begin{equation}
		\mathcal{P}^2 X = \sum_{\varepsilon, \varepsilon'} \Pi_{\varepsilon'}  \Pi_{\varepsilon}  X \Pi_{\varepsilon} \Pi_{\varepsilon'} = \sum_{\varepsilon, \varepsilon'} \delta_{\varepsilon \varepsilon'} \Pi_{\varepsilon}  X \Pi_{\varepsilon} = \sum_{\varepsilon} \Pi_{\varepsilon}  X \Pi_{\varepsilon}  = \mathcal{P} X
	\end{equation}
	and
	\begin{eqnarray}
		\fl
		\Tr X^{\dagger} \mathcal{P} Y = \Tr X^{\dagger}  \sum_{\varepsilon} \Pi_{\varepsilon}  Y \Pi_{\varepsilon}  \nonumber \\
		  = \sum_{\varepsilon}  \Tr \Pi_{\varepsilon}  X^{\dagger}  \Pi_{\varepsilon}  Y   =   \Tr \sum_{\varepsilon}( \Pi_{\varepsilon}  X  \Pi_{\varepsilon})^{\dagger}  Y = \Tr (\mathcal{P} X)^{\dagger}  Y
	\end{eqnarray}
	we obtain property \eref{it:proj}.
	
	\section{Perturbative expansion for effective Hamiltonian}\label{app:proofOfCumulant}
	
	\noindent\textbf{Proof of proposition \ref{prop:effectiveHamExp}.}
		Let us define $ V(\beta) \equiv e^{\beta H_0} e^{- \beta H} $, then it satisfies
		\begin{equation}
			\frac{d}{d \beta} V(\beta) = - \lambda H_{I}(\beta) V(\beta), \qquad V(0) = I,
		\end{equation}
		where $ H_{I}(\beta) $ is defined by \eref{eq:intHamIntPic}. Namely, 
		\begin{eqnarray}
			\fl
			\frac{d}{d \beta} (e^{\beta H_0} e^{- \beta H}) = - e^{\beta H_0} H e^{- \beta H} + e^{\beta H_0} H_0 e^{- \beta H} \nonumber \\
			=  -e^{\beta H_0} (H-H_0) e^{-\beta H_0}  (e^{\beta H_0} e^{- \beta H}) = - \lambda H_{I}(\beta)  (e^{\beta H_0} e^{- \beta H}).
		\end{eqnarray}
		Then representing $ V(\beta) $ by the Dyson series and applying the projector $ \mathcal{P} $ one has
		\begin{equation}\label{eq:PVexpansion}
			\mathcal{P}V(\beta) = I + \sum_{k=1}^{\infty}\lambda^k \mathcal{M}_{k}(\beta)
		\end{equation}
		with $ \mathcal{M}_{k}(\beta) $ defined by \eref{eq:moments}. By the Richter formula \cite[equation (11.1)]{highamFunctionsMatrices2008} one has
		\begin{equation}\label{eq:Richter}
			\log  \mathcal{P} V(\beta) = \int_0^1 (\mathcal{P} V(\beta) - I)(t (\mathcal{P} V(\beta) - I) + I)^{-1}dt
		\end{equation}
		Then we have
		\begin{equation}
			(t (\mathcal{P} V(\beta)  - I) + I)^{-1}  = \sum_{n=0}^{\infty} \lambda^n \sum_{k_1 + \cdots + k_m = n} (-1)^m t^m  \mathcal{M}_{k_1}(\beta) \cdots \mathcal{M}_{k_m}(\beta)
		\end{equation}
		and
		\begin{eqnarray}
			\fl
			(\mathcal{P} V(\beta)  - I)(t (\mathcal{P} V(\beta) - I) + I)^{-1} \nonumber \\ = \sum_{n=1} \lambda^n \sum_{k_0 + \cdots + k_m = n} (-1)^m t^m  \mathcal{M}_{k_0}(\beta) \mathcal{M}_{k_1}(\beta) \cdots \mathcal{M}_{k_m}(\beta)
		\end{eqnarray}
		By substituting it in  \eref{eq:Richter} and taking the integral we have
		\begin{equation}\label{eq:logSeries}
			\log \mathcal{P} V(\beta) = \sum_{n=1}^{\infty} \lambda^n \sum_{k_0 + \cdots + k_m = n} \frac{(-1)^m}{m + 1} \mathcal{M}_{k_0}(\beta) \mathcal{M}_{k_1}(\beta) \cdots \mathcal{M}_{k_m}(\beta).
		\end{equation}
		Taking into account
		\begin{equation}\label{eq:formulaPV}
			\fl
			\mathcal{P} V(\beta)  = \mathcal{P} e^{\beta H_0} e^{- \beta H}  =   \lim\limits_{T \rightarrow + \infty} \frac{1}{T}\int_0^T e^{i H_0 t} e^{\beta H_0} e^{- \beta H}  e^{-i H_0 t} dt = e^{\beta H_0}   \mathcal{P} e^{- \beta H}
		\end{equation}
		we have $ \mathcal{P} e^{- \beta H} = e^{-\beta H_0} \mathcal{P} V(\beta) $. Let us remark that $ H_0 $ commutes with any operator $ \mathcal{P} X $ 
		\begin{equation}\label{eq:commH0P}
			[H_0,  \mathcal{P} X]= \sum_{\varepsilon}( \varepsilon \Pi_{\varepsilon}  X \Pi_{\varepsilon} -  \Pi_{\varepsilon}  X \Pi_{\varepsilon}\varepsilon  ) = 0,
		\end{equation}
		where \eref{eq:PProjRep} was used. Thus, we have
		\begin{equation}
			\mathcal{P} e^{- \beta H} = e^{-\beta H_0} 	e^{\log  \mathcal{P} V(\beta)}  = e^{-\beta (H_0 - \beta^{-1} \log 	\mathcal{P} V(\beta) )} 
		\end{equation}
		and $ \tilde{H} = H_0 - \beta^{-1} \log \mathcal{P} V(\beta)  $, which along with  \eref{eq:logSeries} leads to  \eref{eq:effectiveHamExp}.

	\section{Eigenprojector expansion}\label{app:proofOfExplicit}
	
	\begin{lemma}\label{lemma:auxiliary}
	The following formula holds
		\begin{eqnarray}
			\fl
			\int_{0}^{\beta} d \beta_1  \ldots  \int_{0}^{\beta_{n-1}} d \beta_n e^{ - \sum_{j=1}^n \beta_j \omega_j} \nonumber \\
			= \frac{1}{\prod_{k=1}^n \sum_{j=1}^k\omega_j}
			+ \sum _{p=1}^n (-1)^p \frac{ e^{-\beta  \sum_{i=1}^p \omega_i}}{\left(\prod_{m=1}^p	\sum_{i=m}^p \omega_i\right) \left(\prod_{k=p+1}^n \sum _{j=p+1}^k \omega_j\right)}.
			\label{eq:auxiliarylemma}
		\end{eqnarray}
	\end{lemma}
	
	\noindent \textbf{Proof.}
	Let us denote 
		\begin{equation}
			h_{n}(\beta; \omega_1, \ldots, \omega_n) =	\int_{0}^{\beta} d \beta_1  \ldots  \int_{0}^{\beta_{n-1}} d \beta_n e^{ - \sum_{j=1}^n \beta_j \omega_j},
		\end{equation}
	then, by direct computation, we have
		\begin{eqnarray}
			\fl
			h_{n+1}(\beta; \omega_1, \ldots, \omega_{n+1}) = \int_0^{\beta} d \beta_1 e^{- \omega_1 \beta_1} h_{n}(\beta; \omega_{2}, \ldots, \omega_{n+1}) 
			\nonumber \\
			\fl
			=  \frac{1}{\prod_{k=1}^n \sum _{j=1}^k \omega_{j+1}} \int_{0}^{\beta} e^{- \omega_1 \beta_1}  d \beta_1 
			\nonumber \\
			\fl
			+ \sum_{p=1}^n (-1)^p \frac{1}{\left(\prod_{m=1}^p \sum_{i=m}^p \omega_{i+1}\right) \left(\prod_{k=p+1}^n \sum _{j=p+1}^k \omega_{j+1}\right)}  \int_{0}^{\beta} e^{- \omega_1 \beta_1} e^{-\beta_1  \sum_{i=1}^p \omega_{i+1}}  d \beta_1  
			\nonumber \\
			\fl
			=  \frac{1}{\prod_{k=2}^{n+1}\sum _{j=2}^k \omega_{j}} \frac{1 -e^{- \omega_1 \beta_1}}{\omega_1}
			\nonumber \\
			\fl
			+ \sum_{p=1}^n (-1)^p \frac{1}{\left(\prod_{m=2}^{p+1} \sum_{i=m}^{p+1} \omega_{i}\right) \left(\prod_{k=p+2}^{n+1} \sum _{j=p+2}^k \omega_{j}\right)}  \frac{1 - e^{- \beta \sum_{i=1}^{p+1} \omega_{i}}}{ \sum_{i=1}^{p+1} \omega_{i}}
			\nonumber \\
			\fl
			=  \frac{1}{\prod_{k=2}^{n+1}\sum _{j=2}^k \omega_{j}} \frac{1 -e^{- \omega_1 \beta_1}}{\omega_1}
			\nonumber \\
			\fl
			- \sum_{p=2}^{n+1} (-1)^p \frac{1}{\left(\prod_{m=2}^{p} \sum_{i=m}^{p} \omega_{i}\right) \left(\prod_{k=p+1}^{n+1} \sum _{j=p+1}^k \omega_{j}\right)}  \frac{1 - e^{- \beta \sum_{i=1}^{p} \omega_{i}}}{ \sum_{i=1}^{p} \omega_{i}}
			\nonumber \\
			\fl
			=  \frac{1}{\prod_{k=2}^{n+1}\sum _{j=2}^k \omega_{j}} \frac{1}{\omega_1} - \sum_{p=2}^{n+1} (-1)^p \frac{1}{\left(\prod_{m=1}^{p} \sum_{i=m}^{p} \omega_{i}\right) \left(\prod_{k=p+1}^{n+1} \sum _{j=p+1}^k \omega_{j}\right)} 
			\nonumber \\
			\fl
			- \frac{1}{\prod_{k=2}^{n+1}\sum _{j=2}^k \omega_{j}} \frac{e^{-\beta \omega_1}}{\omega_1} 
			+ \sum_{p=2}^{n+1} (-1)^p \frac{1}{\left(\prod_{m=1}^{p} \sum_{i=m}^{p} \omega_{i}\right) \left(\prod_{k=p+1}^{n+1} \sum _{j=p+1}^k \omega_{j}\right)} e^{- \beta \sum_{i=1}^{p} \omega_{i}}
			\nonumber \\
			\fl
			=  - \sum_{p=1}^{n+1} (-1)^p \frac{1}{\left(\prod_{m=1}^{p} \sum_{i=m}^{p} \omega_{i}\right) \left(\prod_{k=p+1}^{n+1} \sum _{j=p+1}^k \omega_{j}\right)}
			\nonumber \\
			\fl
			+ \sum_{p=1}^{n+1} (-1)^p \frac{1}{\left(\prod_{m=1}^{p} \sum_{i=m}^{p} \omega_{i}\right) \left(\prod_{k=p+1}^{n+1} \sum _{j=p+1}^k \omega_{j}\right)} e^{- \beta \sum_{i=1}^{p} \omega_{i}}.
			\label{eq:longCalc}
		\end{eqnarray}
		Using
		\begin{equation}
			h_{n}(0; \omega_1, \ldots, \omega_n) = 0 
		\end{equation}
		we have
		\begin{equation}
			\sum_{p=1}^{n} (-1)^p \frac{1}{\left(\prod_{m=1}^p \sum_{i=m}^p \omega_i\right) \left(\prod_{k=p+1}^{n} \sum _{j=p+1}^k \omega_j\right)} = -\frac{1}{\prod_{k=1}^n \sum_{j=1}^k\omega_j},
		\end{equation}
		then
		\begin{eqnarray}
			\fl
			\sum_{p=1}^{n+1} (-1)^p \frac{1}{\left(\prod_{m=1}^p \sum_{i=m}^p \omega_i\right) \left(\prod_{k=p+1}^{n+1} \sum _{j=p+1}^k \omega_j\right)} 
			\nonumber \\
			\fl
			= \sum_{p=1}^{n} (-1)^p \frac{1}{\left(\prod_{m=1}^p \sum_{i=m}^p \omega_i\right) \left(\prod_{k=p+1}^{n} \sum _{j=p+1}^k \omega_j\right)} \frac{1}{\sum _{j=n+1}^{n+1} \omega_j} 
			+(-1)^{n+1} \frac{1}{\left(\prod_{m=1}^{n+1} \sum_{i=m}^{n+1} \omega_i\right) } 
			\nonumber \\
			\fl
			=
			-\frac{1}{\prod_{k=1}^n \sum_{j=1}^k\omega_j} \frac{1}{\omega_{n+1}} 
			+(-1)^{n+1} \frac{1}{\prod_{m=1}^{n} \left(\sum_{i=m}^{n} \omega_i + \omega_{n+1}\right) }  \frac{1}{\omega_{n+1}} 
			= \frac{1}{\prod_{k=1}^{n+1} \sum_{j=1}^k\omega_j}.
		\end{eqnarray}
		Substituting it in \eref{eq:longCalc} we obtain  \eref{eq:auxiliarylemma}.
		
	
	\begin{lemma}\label{lemma:aver}
		The following formula holds
		\begin{eqnarray}
			\fl
			\int_{0}^{\beta} d \beta_1  \ldots  \int_{0}^{\beta_{n-1}} d \beta_n e^{ - \sum_{j=1}^{n-1} \beta_j \omega_j + \beta_n \sum_{j=1}^{n-1} \omega_j } = \nonumber\\
			\frac{1}{\prod _{k=1}^{n-1} \sum _{j=1}^k \omega_j} \left(\beta - \sum _{k=1}^{n-1} \frac{1}{\sum _{j=1}^k \omega _j} \right) \nonumber\\
			- \sum_{p=1}^{n-1} \frac{(-1)^p }{\left(\prod_{m=2}^p \sum _{i=m}^p \omega _i\right) \left(\sum_{r=1}^p \omega_r\right)^2 
				\left(\prod _{k=p+1}^{n-1} \sum _{j=p+1}^k
				\omega _j\right)} e^{-\beta \sum_{i=1}^p \omega_i}.
			\label{eq:lemmaaver}
		\end{eqnarray}
	\end{lemma}

	
	\noindent \textbf{Proof.} 
		From lemma \ref{lemma:auxiliary} we have
		\begin{eqnarray}
			\fl
				h_{n}(\beta; \omega_1, \ldots, \omega_n)
			= \frac{1}{\sum _{j=1}^n
				\omega_j \prod _{k=1}^{n-1} \sum _{j=1}^k
				\omega_j}
			 \nonumber\\
			 \fl
			+ \sum _{p=1}^{n-1} (-1)^p \frac{ e^{-\beta  \sum_{i=1}^p \omega _i}}{\left(\prod _{m=1}^p
				\sum _{i=m}^p \omega _i\right) \left(\prod_{k=p+1}^n \sum _{j=p+1}^k \omega
				_j\right)}
			+ (-1)^n \frac{ e^{-\beta  \sum_{i=1}^n \omega_i}}{\sum _{i=1}^n \omega_i \left(\prod _{m=2}^n
				\sum _{i=m}^n \omega_i\right) }.
		\end{eqnarray}
		
		Taking the limit
		\begin{equation}
			\sum_{i=1}^n \omega_i \rightarrow 0
		\end{equation}
		we obtain \eref{eq:lemmaaver}.
	
\noindent \textbf{Proof of proposition \ref{prop:EigenOperExp}.}	
		Using expansion \eref{eq:HIEigenOperExp}-\eref{eq:HIEigenOper} we have
		\begin{equation}
			H_I(\beta) =  e^{\beta H_0}  H_I e^{-\beta H_0} =\sum_{\omega} e^{-\beta \omega} D_{\omega}.
		\end{equation}
		Then
		\begin{equation}\label{eq:projectedProdOfOp}
			\mathcal{P} H_I(\beta_1) \cdots H_I(\beta_k) = \sum_{\omega_1, \ldots, \omega_k} e^{- \beta_1 \omega_1 - \ldots - \beta_k \omega_k} \mathcal{P}(D_{\omega_1}  \ldots D_{\omega_k}).
		\end{equation}
		Let us calculate
		\begin{eqnarray}
			\fl
			\mathcal{P}(D_{\omega_1}  \ldots D_{\omega_k}) = \lim\limits_{T \rightarrow + \infty}  \frac{1}{T} \int_0^T  dt e^{i H_0 t} D_{\omega_1}  \ldots D_{\omega_k} e^{-i H_0 t} 
			\nonumber \\
			= \lim\limits_{T \rightarrow + \infty}  \frac{1}{T} \int_0^T  dt \;  e^{i H_0 t} D_{\omega_1}e^{-i H_0 t}  \ldots e^{i H_0 t} D_{\omega_k} e^{-i H_0 t} 
			\nonumber \\
			= \lim\limits_{T \rightarrow + \infty}  \frac{1}{T} \int_0^T  dt \; e^{ i (\omega_1 + \cdots + \omega_k) t} D_{\omega_1} \cdots D_{\omega_k} 
			\nonumber \\
			=  D_{\omega_1} \cdots D_{\omega_{k-1}} D_{-\omega_1 - \ldots - \omega_{k-1}} \delta_{\omega_1 + \cdot + \omega_k, 0} .
		\end{eqnarray}
		Substituting it in \eref{eq:projectedProdOfOp} we have
		\begin{eqnarray}
			\fl
			\mathcal{P} H_I(\beta_1) \cdots H_I(\beta_k) \nonumber\\
			= \sum_{\omega_1, \ldots, \omega_{k-1}} e^{- (\beta_1 - \beta_k) \omega_1 - \ldots - (\beta_{k-1} - \beta_k) \omega_{k-1}}  D_{\omega_1} \cdots D_{\omega_{k-1}} D_{-\omega_1 - \ldots - \omega_{k-1}}.
		\end{eqnarray}
		
		Then by \eref{eq:moments} and lemma \ref{lemma:aver}  we have \eref{en:EigenOperExp}.

	Several first operators $ \mathcal{M}_{k}(\beta) $ are
	\begin{eqnarray}
		\fl
		\mathcal{M}_{1}(\beta) = - \beta D_{0},\\
		\fl\mathcal{M}_{2}(\beta)  = \sum_{\omega}  \frac{\beta  \omega +e^{-\beta \omega }-1}{\omega^2} D_{\omega} D_{-\omega}  = \sum_{\omega \neq 0}  \frac{\beta  \omega +e^{-\beta \omega }-1}{\omega^2} D_{\omega} D_{\omega}^{\dagger} + \frac{\beta}{2} D_{0}^2,\\
		\fl\mathcal{M}_{3}(\beta) = -\sum_{\omega_1, \omega_2}\left(\frac{\beta -\frac{1}{\omega_1}-\frac{1}{\omega _1+\omega _2}}{\omega_1 \left(\omega_1+\omega_2\right)}+\frac{e^{-\beta  \omega_1}}{\omega _1^2 \omega_2}-\frac{e^{-\beta  \left(\omega_1+\omega _2\right)}}{\omega _2	\left(\omega _1+\omega _2\right)^2}\right) D_{\omega_1} D_{\omega_2}  D_{-\omega_1-\omega_2}.
	\end{eqnarray}
	This leads to
	\begin{equation}
		\mathcal{M}_{2}(\beta) - \frac12 (\mathcal{M}_{1}(\beta))^2 =  \sum_{\omega \neq 0}  \frac{\beta  \omega +e^{-\beta \omega }-1}{\omega^2} D_{\omega} D_{\omega}^{\dagger}.
	\end{equation}
	Substituting this expression  and $ \mathcal{M}_{1}(\beta) $ in \eref{eq:firstTerms} leads to \eref{eq:expanToSec}.

	Similarly, higher order cumulants could be calculated, e.g. 
	\begin{eqnarray}
		\fl
		\mathcal{M}_3(\beta) - \frac12 \mathcal{M}_2(\beta) \mathcal{M}_1(\beta) - \frac12 \mathcal{M}_1(\beta) \mathcal{M}_2(\beta) +  \frac13 (\mathcal{M}_1(\beta))^3 
		\nonumber\\
		\fl
		=\sum_{\omega_1, \omega_2}\left(\frac{\beta -\frac{1}{\omega_1}-\frac{1}{\omega _1+\omega _2}}{\omega_1 \left(\omega_1+\omega_2\right)}+\frac{e^{-\beta  \omega_1}}{\omega _1^2 \omega_2}-\frac{e^{-\beta  \left(\omega_1+\omega _2\right)}}{\omega _2	\left(\omega _1+\omega _2\right)^2}\right) D_{\omega_1} D_{\omega_2}  D_{-\omega_1-\omega_2} 
		\nonumber\\
		\fl
		+ \beta \frac12 \sum_{\omega}  \frac{\beta  \omega +e^{-\beta \omega }-1}{\omega^2} D_{\omega} D_{-\omega} D_0  + \beta \frac12 \sum_{\omega}  \frac{\beta  \omega +e^{-\beta \omega }-1}{\omega^2}  D_0 D_{\omega} D_{-\omega} - \frac13 \beta^3 D_0^3.
	\end{eqnarray}

	\section{Average of second correction with respect to Gibbs state for free Hamiltonian}\label{app:simplCalc}
	
	Let us express $ \langle D_{\omega}^{\dagger} D_{\omega} \rangle_0  $ in terms of $  \langle D_{\omega}  D_{\omega}^{\dagger}  \rangle_0 $ as
	\begin{eqnarray}
		\fl
		\langle D_{\omega}^{\dagger} D_{\omega} \rangle_0 = \Tr D_{\omega}^{\dagger} D_{\omega} Z^{-1}e^{- \beta H_0} = Z^{-1} \Tr D_{\omega}^{\dagger}  e^{- \beta H_0} e^{\beta H_0} D_{\omega} e^{- \beta H_0} 
		\nonumber \\
		\fl
		= Z^{-1} \Tr D_{\omega}^{\dagger}  e^{- \beta H_0} e^{-\beta \omega} D_{\omega} 
		=  e^{-\beta \omega}  \Tr D_{\omega}  D_{\omega}^{\dagger} Z^{-1}   e^{- \beta H_0} = e^{-\beta \omega} \langle D_{\omega}  D_{\omega}^{\dagger}  \rangle_0.
	\end{eqnarray}
	Taking into account \eref{eq:secOrderHeffTerm} we have
	\begin{eqnarray}
		\fl
		-\langle \tilde{H}^{(2)} \rangle_0 = \frac{\beta}{2} \sum_{\omega \neq 0} f(\beta \omega)  \langle D_{\omega} D_{\omega}^{\dagger} \rangle_0 = \frac{\beta}{2} \sum_{\omega > 0} (f(\beta \omega)  \langle D_{\omega} D_{\omega}^{\dagger} \rangle_0 + f(-\beta \omega)  \langle D_{\omega}^{\dagger} D_{\omega} \rangle_0) \nonumber \\
		=  \beta \sum_{\omega > 0} \frac{1}{2}(f(\beta \omega)  + e^{-\beta \omega} f(-\beta \omega)) \langle D_{\omega} D_{\omega}^{\dagger}  =   \sum_{\omega > 0} \frac{1 - e^{- \beta \omega }}{\omega} \langle D_{\omega} D_{\omega}^{\dagger} \rangle_0. \label{eq:simpCalcOfSecOrdAv}
	\end{eqnarray}
	Similarly, taking into account \eref{en:monotoneSecOrder}  we have 
	\begin{eqnarray}
		\fl
		-\langle \partial_{\beta} \tilde{H}^{(2)} \rangle_0 = \frac{\beta}{2} \sum_{\omega \neq 0} f_1(\beta \omega)  \langle D_{\omega} D_{\omega}^{\dagger} \rangle_0 
		=   \sum_{\omega > 0} \frac{1}{2}(f_1(\beta \omega)  + e^{-\beta \omega} f_1(-\beta \omega)) \langle D_{\omega} D_{\omega}^{\dagger} \rangle_0  \nonumber \\
		= \sum_{\omega > 0} \frac{1 - e^{- \beta \omega }}{\beta \omega} \langle D_{\omega} D_{\omega}^{\dagger} \rangle_0 = \frac{1}{\beta}\langle \tilde{H}^{(2)} \rangle_0.
	\end{eqnarray}
	
	\section{Perturbative expansion of mean force Hamiltonian for effective Gibbs state}\label{sec:meanForceEffGibbs}

	\textbf{Proof of proposition \ref{prop:meanForce}.}
		Taking into account  \eref{eq:formulaPV} we have
		\begin{eqnarray}
			\fl
			\Tr_B \mathcal{P} e^{- \beta H}  = \Tr_B e^{ -\beta H_0} \mathcal{P}V(\beta) \nonumber \\
			= e^{ -\beta H_S} \Tr_B e^{ -\beta H_B} \mathcal{P}V(\beta)  
			= e^{ -\beta H_S} \Tr_B\mathcal{P}V(\beta)  e^{ -\beta H_B} . \label{eq:fromulaTrBP}
		\end{eqnarray}
		Due to  \eref{eq:commH0P} it can also be written as  
		\begin{eqnarray}
			\Tr_B \mathcal{P} e^{- \beta H}  = \Tr_B \mathcal{P}V(\beta) e^{ -\beta H_0}  = ( \Tr_B \mathcal{P}V(\beta) e^{ -\beta H_B}) e^{ -\beta H_S},
		\end{eqnarray}
		so
		\begin{equation}
			[ \Tr_B \mathcal{P}V(\beta) e^{ -\beta H_B}, e^{ -\beta H_S}] = 0.
		\end{equation}
		By \eref{eq:PVexpansion} we have
		\begin{eqnarray}
			\fl
			\Tr_B \mathcal{P}V(\beta) e^{ -\beta H_B} = \Tr_B e^{- \beta H_B} \left( \sum_{k=0}^{\infty}\lambda^k \mathcal{P}  \mathcal{M}_{k}(\beta)\right)  \nonumber \\
			= Z_B \left( 1 + \sum_{k=1}^{\infty}\lambda^k \Tr_B Z_B^{-1} e^{- \beta H_B} \mathcal{M}_{k}(\beta)\right) = Z_B \sum_{k=1}^{\infty}\lambda^k  \langle   \mathcal{M}_{k}(\beta) \rangle_{B}.
		\end{eqnarray}
		Taking into account \eref{eq:fromulaTrBP} we have 
		\begin{equation}
			Z_B^{-1}e^{-\beta \tilde{H}_{\rm mf}} = Z_B^{-1}	\Tr_B \mathcal{P} e^{- \beta H}  = e^{- \beta H_S} \sum_{k=1}^{\infty}\lambda^k  \langle   \mathcal{M}_{k}(\beta) \rangle_{B}.
		\end{equation}
		Then the proof follows the proof of proposition \ref{prop:effectiveHamExp} (see  \ref{app:proofOfCumulant}) replacing $  \mathcal{M}_{k}(\beta) $ with $ \langle   \mathcal{M}_{k}(\beta) \rangle_{B} $ and $ H_0 $ with $ H_S $.

	Similarly to \eref{eq:firstTerms} the first several terms are
	\begin{equation}
		\fl
		\tilde{H}_{\rm mf} = H_S -   \lambda \beta^{-1} \langle \mathcal{M}_{1}(\beta) \rangle_{B}  - \beta^{-1} \lambda^2 \left( \langle \mathcal{M}_{2}(\beta) \rangle_{B}- \frac12 (\langle \mathcal{M}_{1}(\beta) \rangle_{B})^2\right)  + O(\lambda^3)
	\end{equation}
	or using proposition \ref{prop:EigenOperExp} similarly to \eref{eq:expanToSec} we have \eref{eq:meanForcSecOrder}.

	\section{Calculation of mean force Hamiltonian}
	\label{app:meanForcSecOrderExplicit}
	Due to \eref{eq:subEigenProj} we have
	\begin{equation}\label{eq:Domega}
		D_{\omega} = \sum_{\omega_1} A_{\omega - \omega_1} \otimes B_{\omega_1}= \sum_{\omega_1} A_{\omega_1-\omega}^{\dagger} \otimes B_{\omega_1} = \sum_{\omega_1} A_{\omega_1}^{\dagger} \otimes B_{\omega_1 + \omega},
	\end{equation}
	 then
	\begin{equation}\label{eq:DomegaDomegaDagger}
		D_{\omega} 	D_{\omega}^{\dagger} = \sum_{\omega_1, \omega_2} A_{ \omega_1 }^{\dagger}  A_{\omega_2}  \otimes B_{\omega_1 + \omega} B_{\omega_2 + \omega}^{\dagger} .
	\end{equation}
	The second equation of \eref{eq:subEigenProj} also leads to $ 	e^{\beta H_B} B_{\omega}	e^{- \beta H_B} =  e^{\beta [H_B, \; \cdot \;]} B_{\omega}  = 	e^{- \beta \omega} B_{\omega} $, then $ B_{\omega}	e^{- \beta H_B} = 	e^{- \beta \omega} e^{- \beta H_B} B_{\omega} $. Applying trace to both sides of this equation we have $ (1-e^{- \beta \omega} )\Tr_B B_{\omega}	e^{- \beta H_B} = 0 $. Thus, we have
	\begin{equation}
		\langle  B_{\omega} \rangle_{B} = \langle  B_{0} \rangle_{B} \delta_{\omega,0}
	\end{equation}
	\begin{equation}
		\langle  B_{\omega_1} B_{\omega_2}^{\dagger} \rangle_{B} = \langle  B_{\omega_1} B_{\omega_1}^{\dagger} \rangle_{B} \delta_{\omega_1,\omega_2}.
	\end{equation}
	Then \eref{eq:Domega} and \eref{eq:DomegaDomegaDagger} take the form
	\begin{equation}
		\langle D_{0}  \rangle_B = \sum_{\omega_1} A_{ - \omega_1}  \langle B_{\omega_1} \rangle_B =  A_{0}  \langle B_{0} \rangle_B
	\end{equation}
	and
	\begin{eqnarray}
		\fl
		\langle  D_{\omega} D_{\omega}^{\dagger} \rangle_{B} =   \sum_{\omega_1, \omega_2} A_{ \omega_1 }^{\dagger}  A_{\omega_2}  \langle  B_{\omega_1 + \omega} B_{\omega_2 + \omega}^{\dagger} \rangle_B =  \sum_{\omega_1} A_{ \omega_1 }^{\dagger}  A_{\omega_1}  \langle  B_{\omega_1 + \omega} B_{\omega_1 + \omega}^{\dagger} \rangle_B \nonumber\\
		=  \sum_{\omega_1 \neq 0} A_{ \omega_1 }^{\dagger}  A_{\omega_1}  \langle  B_{\omega_1 + \omega} B_{\omega_1 + \omega}^{\dagger} \rangle_B + A_{0}^{2}   \langle  B_{\omega} B_{ \omega}^{\dagger} \rangle_B.
	\end{eqnarray}
	Hence,  after substituting these formulae into \eref{eq:meanForcSecOrder} we have
	\begin{eqnarray}
		\fl
		\tilde{H}_{\rm mf} = H_S + \lambda   \langle  B_{0} \rangle_{B} A_{0}  
		- \lambda^2 \frac{\beta}{2 } \biggl((\sum_{\omega \neq 0, \omega_1 \neq 0}   f(\beta \omega )   \langle  B_{\omega_1 + \omega} B_{\omega_1 + \omega}^{\dagger} \rangle_B  + \langle  B_{\omega_1} B_{\omega_1}^{\dagger}  \rangle_B ) A_{ \omega_1 }^{\dagger}  A_{\omega_1}	\nonumber \\
		+   A_{0}^2  ( \sum_{\omega \neq 0}  f(\beta \omega ) \langle  B_{\omega} B_{ \omega}^{\dagger} \rangle_B +\langle  B_{0}^2  \rangle - \langle  B_{0} \rangle_B^2) \biggr)  + O(\lambda^3).
	\end{eqnarray}
	Assuming by continuity $ f(0) = 1 $ this equation reduces to \eref{eq:meanForcSecOrderExplicit}.

	\section{Calculations for the examples}\label{app:calcForExamp}
	
	We provide fewer details for the second and  the third example because they are fully analogous to the first one.
	
	\subsection{Two two-level systems}
	
	1) For the off-resonance case we have
	\begin{equation}
		H_0 = \omega_a \sigma_a^+\sigma_a^- + \omega_b \sigma_b^+\sigma_b^-, \qquad H_I =( \sigma_a^- + \sigma_a^+)  (g^* \sigma_b^-  + g \sigma_b^+).
	\end{equation}
	As $ [\omega_i \sigma_i^+\sigma_i^-, \sigma_i^{\pm}] = - (\mp \omega_i)  \sigma_i^{\pm}$ for $ i = a,b $, then
	\begin{equation}
		D_{\omega_a - \omega_b} = D_{\omega_b - \omega_a}^{\dagger}= g \sigma_a^- \sigma_b^+, \qquad D_{\omega_a + \omega_b} =  D_{-(\omega_a + \omega_b)}^{\dagger}= g^* \sigma_a^- \sigma_b^-.
	\end{equation}
	As $ n_i = \sigma_i^+ \sigma_i^- = 1- \sigma_i^- \sigma_i^+ $ for $ i = a,b $, then
	\begin{eqnarray}
		D_{\omega_a - \omega_b} D_{\omega_a - \omega_b}^{\dagger} &= |g|^2 \sigma_a^- \sigma_a^+ \sigma_b^+ \sigma_b^- = |g|^2 (1-n_a) n_b,\\
		D_{\omega_a - \omega_b}^{\dagger} D_{\omega_a - \omega_b} &= |g|^2 \sigma_a^+ \sigma_a^- \sigma_b^- \sigma_b^+ = |g|^2 n_a (1-n_b),\\
		D_{\omega_a + \omega_b} D_{\omega_a +\omega_b}^{\dagger} &= |g|^2 \sigma_a^- \sigma_a^+ \sigma_b^- \sigma_b^+ = |g|^2 (1-n_a) (1-n_b), \\
		D_{-\omega_a - \omega_b} D_{-\omega_a -\omega_b}^{\dagger} &= |g|^2 \sigma_a^+ \sigma_a^- \sigma_b^+ \sigma_b^- = |g|^2 n_a n_b.
	\end{eqnarray}
	Substituting it in \eref{eq:expanToSec} we obtain \eref{eq:twoTwoLevelEff}.
	
	As
	\begin{equation}
		\langle n_i \rangle_0 \equiv \frac{\Tr n_i e^{- \beta \omega_i n_i}}{\Tr  e^{- \beta \omega_i n_i}} =  \frac{1}{e^{\beta \omega_i} + 1}
	\end{equation}
	for $ i = a,b $, then by  \eref{eq:simpCalcOfSecOrdAv} we have
	\begin{eqnarray}
		\fl
		-\langle H^{(2)} \rangle_0 =  |g|^2 \biggl( \frac{1 - e^{- \beta (\omega_a - \omega_b) }}{ \omega_a - \omega_b}  (1-\langle n_a \rangle_0) \langle n_b \rangle_0 + \frac{1 - e^{- \beta (\omega_a + \omega_b) }}{ \omega_a + \omega_b}(1 -\langle n_a \rangle_0) (1 -\langle n_b \rangle_0 )\biggr) \nonumber\\
		= |g|^2 \frac{\omega_a \tanh \frac{\beta \omega_a}{2} - \omega_b \tanh \frac{\beta  \omega_b}{2}}{\omega_a^2 - \omega_b^2}.
	\end{eqnarray}
	Thus, by  \eref{eq:deltaEntropySecOrdSimp} we obtain  \eref{eq:offResEnt}.
	
	2) For the resonance case we have
	\begin{equation}
		H_0 = \omega_a (\sigma_a^+\sigma_a^- + \sigma_b^+\sigma_b^-), \qquad H_I =( \sigma_a^- + \sigma_a^+)  (g^* \sigma_b^-  + g \sigma_b^+) +  \delta \omega \sigma_b^+\sigma_b^-.
	\end{equation}
	Now the terms analogous to $ D_{\omega_a - \omega_b}$ and $ D_{\omega_b - \omega_a} $ contribute to $ D_0 $
	\begin{equation}
		\fl
		D_{0} = D_{0}^{\dagger}= (g \sigma_a^- \sigma_b^+ + g^* \sigma_a^+ \sigma_b^- ) +  \delta \omega\sigma_b^+\sigma_b^-, \qquad D_{\omega_a + \omega_b} =  D_{-(\omega_a + \omega_b)}^{\dagger}= g^* \sigma_a^- \sigma_b^-.
	\end{equation}
	Substituting it in \eref{eq:expanToSec} we obtain \eref{eq:twoTwoLevelEffRes}.
	
	As
	\begin{equation}
		\langle n_i \rangle_0=  \frac{1}{e^{\beta \omega_a} + 1},
	\end{equation}
	we have
	\begin{equation}
		-\langle H^{(2)} \rangle_0 =  |g|^2  \frac{1 - e^{- 2 \beta \omega_a }}{ 2 \omega_a}(1 -\langle n_a \rangle_0) (1 -\langle n_b \rangle_0 ) = |g|^2 \frac{\tanh \frac{\omega_a \beta}{2}}{2 \omega_a}.
	\end{equation}
	Thus, by  \eref{eq:deltaEntropySecOrdSimp} we obtain  \eref{eq:resEnt}.
	
	\subsection{Two oscillators}
	
	1) For the off-resonance case we have
	\begin{equation}
		H_0 = \omega_a a^{\dagger} a+ \omega_b b^{\dagger} b, \qquad H_I =( a + a^{\dagger})  (g^* b  + g  b^{\dagger}),
	\end{equation}
	\begin{equation}
		D_{\omega_a - \omega_b} = D_{\omega_b - \omega_a}^{\dagger}  = g a b^{\dagger}, \qquad D_{\omega_a + \omega_b} = D_{-(\omega_a + \omega_b)}^{\dagger} = g^* a b\\
	\end{equation}
	and
	\begin{eqnarray}
		D_{\omega_a - \omega_b} D_{\omega_a - \omega_b}^{\dagger} &= |g|^2 a a^{\dagger} b^{\dagger} b = |g|^2 (n_a + 1) n_b,\\
		D_{\omega_a + \omega_b} D_{\omega_a + \omega_b}^{\dagger} &= |g|^2 aa^{\dagger}  b  b^{\dagger} = |g|^2 (n_a + 1) (n_b + 1),\\
		D_{\omega_b - \omega_a} D_{\omega_b - \omega_a}^{\dagger} &= |g|^2 a^{\dagger} a b b^{\dagger} = |g|^2 n_a (n_b +1 ),\\
		D_{-(\omega_a + \omega_b)} D_{-(\omega_a + \omega_b)}^{\dagger} &= |g|^2 a^{\dagger}a  b^{\dagger} b = |g|^2 n_a n_b .
	\end{eqnarray}
	Substituting it in \eref{eq:expanToSec} we obtain \eref{eq:twoOscEff}.
	
	As
	\begin{equation}
		\langle n_i \rangle_0 =  \frac{1}{e^{\beta \omega_i} - 1}
	\end{equation}
	for $ i = a,b $, then by  \eref{eq:simpCalcOfSecOrdAv} we have
	\begin{eqnarray}
		\fl
		-\langle H^{(2)} \rangle_0 =  |g|^2 \biggl( \frac{1 - e^{- \beta (\omega_a - \omega_b) }}{ \omega_a - \omega_b}  (1+\langle n_a \rangle_0) \langle n_b \rangle_0 + \frac{1 - e^{- \beta (\omega_a + \omega_b) }}{ \omega_a + \omega_b}(1 +\langle n_a \rangle_0) (1 +\langle n_b \rangle_0 )\biggr) \nonumber\\
		 = |g|^2 \frac{\omega_a  \coth \frac{\beta \omega_b }{2} - \omega_b \coth \frac{\beta  \omega_a}{2}}{\omega_a^2 - \omega_b^2}.
	\end{eqnarray}
	By  \eref{eq:deltaEntropySecOrdSimp} we obtain  \eref{eq:offResEntOsc}.
	
	2)  For resonance case \eref{eq:twoTwoLevel} we have
	\begin{equation}
		H_0 = \omega_a (a^{\dagger} a + b^{\dagger} b), \qquad H_I =( a + a^{\dagger})  (g^* b  + g  b^{\dagger}) +  \delta \omega b^{\dagger} b,
	\end{equation}
	\begin{equation}
		D_{0} = D_{0}^{\dagger}= (g  a b^{\dagger} + g^* a^{\dagger} b ) +  \delta \omega b^{\dagger} b, \quad D_{\omega_a + \omega_b} =  D_{-(\omega_a + \omega_b)}^{\dagger}= g^* a b.
	\end{equation}
	Substituting it in \eref{eq:expanToSec} we obtain \eref{eq:twoOscEffRes}.
	
	As
	\begin{equation}
		\langle n_i \rangle_0=  \frac{1}{e^{\beta \omega_a} - 1},
	\end{equation}
	we have
	\begin{equation}
		-\langle H^{(2)} \rangle_0 =  |g|^2  \frac{1 - e^{- 2 \beta \omega_a }}{ 2 \omega_a}(1 +\langle n_a \rangle_0) (1 +\langle n_b \rangle_0 ) =  |g|^2 \frac{\coth \frac{\omega_a \beta}{2}}{2 \omega_a}.
	\end{equation}
	Thus, by  \eref{eq:deltaEntropySecOrdSimp} we obtain  \eref{eq:resEntOsc}.
	
	\subsection{Two-level system and oscillator}

	1) For the off-resonance case we have
	\begin{equation}
		H_0 = \omega_a \sigma_+\sigma_-+ \omega_b b^{\dagger} b, \qquad H_I =( a + a^{\dagger})  (g^* b  + g  b^{\dagger}),
	\end{equation}
	\begin{equation}
		D_{\omega_a - \omega_b} = D_{\omega_b - \omega_a}^{\dagger}  = g \sigma_- b^{\dagger}, \qquad D_{\omega_a + \omega_b} = D_{-(\omega_a + \omega_b)}^{\dagger} = g^* \sigma_- b\\
	\end{equation}
	and
	\begin{eqnarray}
		D_{\omega_a - \omega_b} D_{\omega_a - \omega_b}^{\dagger} &= |g|^2 \sigma_- \sigma_+ b^{\dagger} b = |g|^2 (1-n_a) n_b,\\
		D_{\omega_a + \omega_b} D_{\omega_a + \omega_b}^{\dagger} &= |g|^2 \sigma_- \sigma_+  b  b^{\dagger} = |g|^2 (1-n_a) (n_b + 1),\\
		D_{\omega_b - \omega_a} D_{\omega_b - \omega_a}^{\dagger} &= |g|^2 \sigma_+  \sigma_- b b^{\dagger} = |g|^2 n_a (n_b +1 ),\\
		D_{-(\omega_a + \omega_b)} D_{-(\omega_a + \omega_b)}^{\dagger} &= |g|^2 \sigma_+ \sigma_-  b^{\dagger} b = |g|^2 n_a n_b .
	\end{eqnarray}
	As
	\begin{equation}
		\langle n_a \rangle_0 =  \frac{1}{e^{\beta \omega_a} + 1}, \qquad \langle n_b \rangle_0 =  \frac{1}{e^{\beta \omega_b} - 1}
	\end{equation}
	for $ i = a,b $, then by  \eref{eq:simpCalcOfSecOrdAv} we have
	\begin{eqnarray}
		\fl
		-\langle H^{(2)} \rangle_0 =  |g|^2 \biggl( \frac{1 - e^{- \beta (\omega_a - \omega_b) }}{ \omega_a - \omega_b}  (1-\langle n_a \rangle_0) \langle n_b \rangle_0 + \frac{1 - e^{- \beta (\omega_a + \omega_b) }}{ \omega_a + \omega_b}(1 -\langle n_a \rangle_0) (1 +\langle n_b \rangle_0 )\biggr) \nonumber\\
		 = |g|^2 \frac{\omega_a  \tanh \frac{\beta  \omega_a}{2}  \coth \frac{\beta \omega_b }{2}- \omega_b }{\omega_a^2 - \omega_b^2}.
	\end{eqnarray}
	By  \eref{eq:deltaEntropySecOrdSimp} we obtain  \eref{eq:offResEntTLSOsc}.
	
	2)  For resonance case \eref{eq:twoTwoLevel} we have
	\begin{equation}
		H_0 = \omega_a (\sigma_+ \sigma_- + b^{\dagger} b), \qquad H_I =( \sigma_- + \sigma_+)  (g^* b  + g  b^{\dagger}) +  \delta \omega b^{\dagger} b,
	\end{equation}
	\begin{equation}
		D_{0} = D_{0}^{\dagger}= (g  \sigma_- b^{\dagger} + g^* \sigma_+ b ) +  \delta \omega b^{\dagger} b, \quad D_{\omega_a + \omega_b} =  D_{-(\omega_a + \omega_b)}^{\dagger}= g^* \sigma_- b.
	\end{equation}
	As
	\begin{equation}
		\langle n_a \rangle_0 =  \frac{1}{e^{\beta \omega_a} + 1}, \qquad \langle n_b \rangle_0 =  \frac{1}{e^{\beta \omega_a} - 1},
	\end{equation}
	we have
	\begin{equation}
		-\langle H^{(2)} \rangle_0 =  |g|^2  \frac{1 - e^{- 2 \beta \omega_a }}{ 2 \omega_a}(1 - \langle n_a \rangle_0) (1 +\langle n_b \rangle_0 ) =   \frac{|g|^2}{2 \omega_a}.
	\end{equation}
	Thus, by  \eref{eq:deltaEntropySecOrdSimp} we obtain  \eref{eq:resEntTLSOsc}.
	
	\section*{References}
	
	\bibliographystyle{iopart-num}
	\bibliography{effective}

\end{document}